\documentclass[conference]{IEEEtran}
\usepackage{cite}
\usepackage{amsmath,amssymb,amsfonts}
\usepackage{algorithmic}
\usepackage{graphicx}
\usepackage{textcomp}
\usepackage{xcolor}
\usepackage{fancyhdr}
\usepackage{etoolbox}
\usepackage{comment}
\usepackage{microtype}
\usepackage{tabularx}
\usepackage[hyphens]{url}

\def\BibTeX{{\rm B\kern-.05em{\sc i\kern-.025em b}\kern-.08em
    T\kern-.1667em\lower.7ex\hbox{E}\kern-.125emX}}

\title{RealityCheck: Bringing Modularity, Hierarchy, and Abstraction to Automated Microarchitectural Memory Consistency Verification} 
\author{Yatin A. Manerkar\quad Daniel Lustig$^*$\quad Margaret Martonosi\\\vspace{2pt}
Princeton University\quad\quad\quad\quad\quad NVIDIA${}^*$\\\vspace{2pt}
\{manerkar,mrm\}@princeton.edu\quad{dlustig}@nvidia.com}

\begin{document}
\maketitle

\newtoggle{inclComments}
\toggletrue{inclComments}

\newtoggle{hlChanges}
\togglefalse{hlChanges}

\newcommand{\acomm}[1]{\iftoggle{inclComments}{\textcolor{red}{(\textbf{#1})}}{}}
\newcommand{\hlc}[1]{\iftoggle{hlChanges}{\hl{#1}}{#1}}
\newcommand{\myverbatim}[1]{\begin{small}\begin{verbatim}#1\end{verbatim}\end{small}}

\pagestyle{plain}


\begin{abstract}

Modern SoCs are heterogeneous parallel systems comprised of components developed by distinct teams and possibly even different vendors. The memory consistency model (MCM) of processors in such SoCs specifies the ordering rules which constrain the values that can be read by load instructions in parallel programs running on such systems. The implementation of required MCM orderings can span components which may be designed and implemented by many different teams. Ideally, each team would be able to specify the orderings enforced by their components independently and then connect them together when conducting MCM verification. However, no prior automated approach for formal hardware MCM verification provided this.

To bring automated hardware MCM verification in line with the realities of the design process, we present RealityCheck, a methodology and tool for automated formal MCM verification of modular microarchitectural ordering specifications.
RealityCheck allows users to specify their designs as a hierarchy of distinct modules connected to each other rather than a single flat specification. It can then automatically verify litmus test programs against these modular specifications. RealityCheck also provides support for abstraction, which enables scalable verification by breaking up the verification of the entire design into smaller verification problems.
We present results for verifying litmus tests on 7 different designs using RealityCheck. These include in-order and out-of-order pipelines, a non-blocking cache, and a heterogeneous processor. Our case studies cover the TSO and RISC-V (RVWMO) weak memory models. RealityCheck is capable of verifying 98 RVWMO litmus tests in under 4 minutes each, and its capability for abstraction enables up to a 32.1\% reduction in litmus test verification time for RVWMO.

\end{abstract}

\section{Introduction}
\label{sec:introduction}

\begin{figure*}[t]
  \centering
  \includegraphics[width=0.9\textwidth]{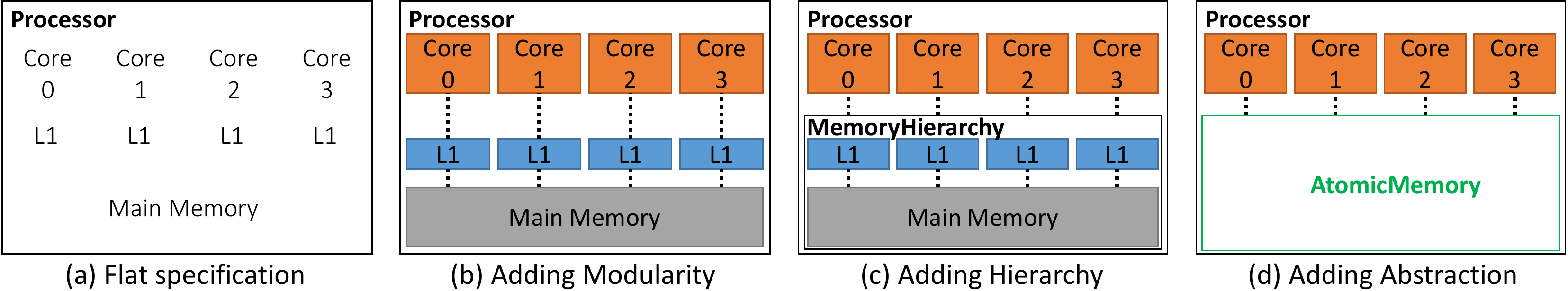}
  \vspace{-10pt}
  \caption{Illustration of adding modularity, hierarchy, and abstraction to a flat design specification.}
  \label{fig:relcheck_pillars}
  \vspace{-10pt}
\end{figure*}

Today's systems-on-chip (SoCs) are complex heterogeneous systems developed by many individuals. Hardware development of the processors used in these SoCs is divided up among different teams, with each team responsible for one or a few components. For instance, one team may be responsible for the pipeline, another for the store buffer, a third for the L1 caches, and a fourth for an accelerator. Each team will have detailed knowledge about the components they design, but may know little about other components. Nevertheless, a processor or SoC created by the interconnection of various components developed by different teams must function correctly.

One measure of the correctness of a processor is its conformance to its memory consistency model (MCM). MCMs consist of rules which constrain the values that can be read by load instructions in parallel programs~\cite{adve:shared}. An ISA-level MCM like that of x86~\cite{intel:x86} serves as both a target for compilers as well as a specification that microarchitectures must implement. If a microarchitecture does not obey the MCM of its ISA, then parallel code compiled to target that ISA will not run correctly on the hardware. MCM bugs have been discovered in hardware in recent years~\cite{haswellTSX,skylakeTSX,manerkar:rtlcheck,elver:tso-ccmanual,lustig:pipecheck},
and will continue to increase as designs become more complex and parallelism becomes ubiquitous.
This makes microarchitectural MCM verification critical to parallel system correctness. There has been much work in recent years on formal MCM specifications of hardware ISAs~\cite{owens:better,sarkar:understanding,pulte:arm,trippel:tricheck}, as well as work on automatically formally verifying that hardware implementations correctly implement those MCMs~\cite{lustig:pipecheck,manerkar:ccicheck,lustig:coatcheck,trippel:tricheck,manerkar:rtlcheck,manerkar:pipeproof}.
However, all of these prior automated approaches suffer from deficiencies which put them at odds with the realities of the processor design cycle. Figure~\ref{fig:relcheck_pillars} shows a graphical depiction of some of these deficiencies. Prior automated formal approaches provide \textit{flat} verification (Figure~\ref{fig:relcheck_pillars}a). There is no way for users to encapsulate component functionality into a unit whose properties only apply to that unit, or to verify a component independently of the rest of the system. In other words, there is no support for \textbf{modularity} as shown in Figure~\ref{fig:relcheck_pillars}b. Similarly, users could not build larger modules from smaller ones as there was no support for \textbf{hierarchy}. Figure~\ref{fig:relcheck_pillars}c shows an example of hierarchy where the L1 and Main Memory modules reside within the \texttt{MemoryHierarchy} module.

Prior work also has no support for \textbf{abstraction}. In other words, there is no way for users to decouple the specification of their component's external behaviour from the specification of its implementation. As an example of abstraction, Figure~\ref{fig:relcheck_pillars}d represents the \texttt{MemoryHierarchy} using an abstract interface \texttt{AtomicMemory}. \texttt{AtomicMemory} specifies the external-facing behaviour of the memory hierarchy, but says nothing about the internal implementation, like how many caches there are. Abstraction facilitates scalable verification by breaking up verification into smaller problems, as this paper shows.

Due to their lack of modularity, prior automated approaches also suffer from a prevalence of \textit{omniscient} or global properties in their design specifications. For example, a property guaranteed by most shared-memory systems today is that of coherence~\cite{sorin:primer}. At an instruction level, coherence requires that there exists a total order on all stores to the same address that is respected by all cores in the system. However, hardware implementations of coherence use \textit{distributed} protocols where each cache (and often a bus/directory) is responsible for enforcing part of the orderings required. None of the hardware components in such an implementation has omniscient visibility of the entire processor, and none of them can make statements about the global behaviour of the system. Thus, global properties such as the coherence definition above reflect a designer's abstraction of the hardware rather than what the hardware is actually doing. If this abstraction is inaccurate, verification using such a specification will be unsound.

To address these deficiencies of prior automated formal hardware MCM verification approaches, we present RealityCheck, a methodology and tool for automated formal MCM verification which supports modularity, hierarchy, and abstraction. RealityCheck allows users to specify their design as a hierarchy of distinct modules connected to each other, closely matching the structure of real hardware implementations. RealityCheck can then automatically verify whether the composition of the various modules exhibits behaviours forbidden by the ISA-level MCM of the processor through bounded verification for suites of litmus test programs. RealityCheck also lets users write interface specifications of the external behaviour of components; it can then verify component implementation specifications against these interfaces. The use of such abstraction allows a design to be verified piece-by-piece, enabling scalable automated microarchitectural MCM verification.

The contributions of this paper are as follows:

\begin{itemize}
    \item \textbf{A Modular Microarchitectural Ordering Specification Language:} RealityCheck develops $\mu$spec++, the first domain-specific language for hierarchical modular specifications of microarchitectural orderings. $\mu$spec++ extends the $\mu$spec domain-specific language from prior work~\cite{lustig:coatcheck} to incorporate modularity, hierarchy, and abstraction.
    \item \textbf{Automated MCM Verification of Modular Design Specifications:} RealityCheck is the first framework for hierarchical modular specification of hardware orderings that enables automated verification of these specifications against ISA-level MCMs for litmus test\footnote{A litmus test is a small program (usually 4-8 instructions long) used to test MCM specifications and implementations.} suites.
    \item \textbf{Scalable Automated Microarchitectural MCM Verification:} RealityCheck is the first automated approach to enable hierarchical MCM verification of a hardware design piece-by-piece. It breaks down verification of the entire design into smaller verification problems.
    \item \textbf{Demonstration:} This paper culminates in the demonstration of RealityCheck's capabilities through its application on multiple modular processor/memory designs. These include both in-order and out-of-order pipelines, a non-blocking cache, and a heterogeneous processor. Our case studies also cover the TSO and RISC-V (RVWMO)~\cite{riscv} weak memory models.
\end{itemize}


\section{Background and Motivation}
\label{sec:background_motiv}

\begin{figure}[t]
  \centering
  
    \renewcommand{\arraystretch}{1.1}
    \centering
    \footnotesize
    \begin{tabular}{|c|c|}
      \hline
      \textbf{Core 0}       & \textbf{Core 1} \\\hline
      (i1) [x] $\leftarrow$ 1  & (i3) [y] $\leftarrow$ 1 \\
      (i2) r1 $\leftarrow$ [y] & (i4) r2 $\leftarrow$ [x] \\\hline
      \multicolumn{2}{|c|}{SC forbids r1=0, r2=0} \\\hline
    \end{tabular}
    \vspace{-5pt}
    \caption{Code for litmus test \texttt{sb}}
    \label{fig:sb_code}
\end{figure}

\begin{figure}[t]
  \centering
  \includegraphics[width=0.25\textwidth]{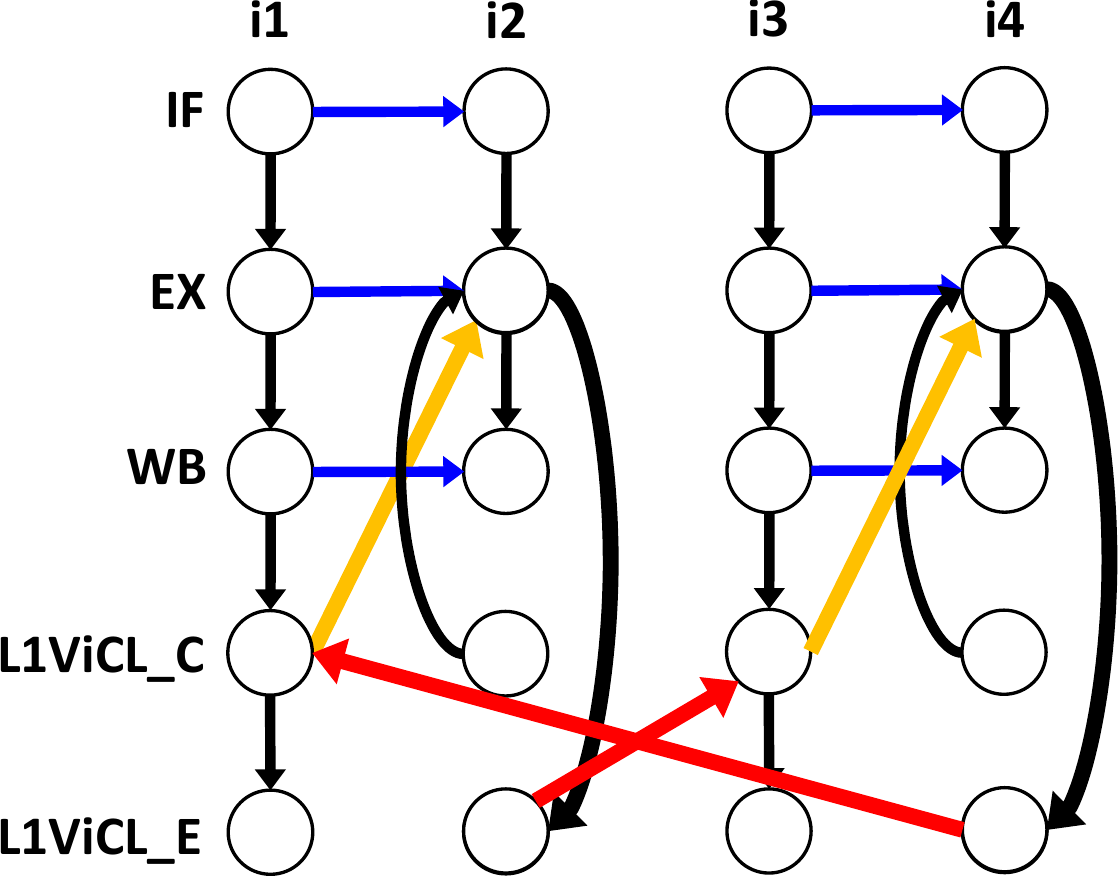}
  \caption{Example $\mu$hb graph for \texttt{sb} litmus test on Figure~\ref{fig:relcheck_pillars}b's processor.}
  \label{fig:sb_uhb}
  \vspace{-10pt}
\end{figure}

\begin{figure}[t]
  \centering
  \begin{small}
  \begin{verbatim}
Axiom "Read_Initial":
forall microop "i", forall microop "j",
 IsAnyRead i /\ DataFromInitialState i /\
 IsAnyWrite j /\ SameAddress i j =>
   AddEdge((i,L1ViCL_E),(j,L1ViCL_C),"").
\end{verbatim}
\end{small}
  \vspace{-10pt}
  \caption{Example $\mu$spec axiom.}
  \vspace{-15pt}
  \label{fig:example_axiom}
\end{figure}

\subsection{Automated Microarchitectural MCM Verification}
\label{sec:check_bkgrnd}

The simplest and most intuitive MCM is sequential consistency (SC)~\cite{lamport:how}. SC defines correct executions as those in which results are consistent with some total order on all memory operations across all cores, in addition to each core performing its operations in program order. To illustrate MCM analysis, consider the store buffering (\texttt{sb}) litmus test in Figure~\ref{fig:sb_code}. 
Initially, both addresses \texttt{x} and \texttt{y} are 0 by convention. Core 0 sets its flag \texttt{x} and reads the value of core 1's flag \texttt{y}. Meanwhile, core 1 sets its flag \texttt{y} and reads the value of core 0's flag \texttt{x}. Under SC, it is forbidden for both loads to return 0, as there is no total order on all memory operations that would allow this.

Consider the execution of \texttt{sb} on the microarchitecture represented by Figure~\ref{fig:relcheck_pillars}b (henceforth called \texttt{exampleProc}). Assume that each core has 3-stage in-order pipelines of Fetch (IF), Execute (EX), and Writeback (WB) stages, and that the processor aims to implement SC. Prior automated formal MCM verification approaches~\cite{lustig:pipecheck,manerkar:ccicheck,lustig:coatcheck,trippel:tricheck,manerkar:rtlcheck,manerkar:pipeproof} modelled microarchitectural executions as $\mu$hb or microarchitectural happens-before graphs~\cite{lustig:pipecheck}, where nodes represent sub-events in the execution of instructions and edges represent happens-before relationships between nodes. Figure~\ref{fig:sb_uhb} shows a $\mu$hb graph depicting the execution of \texttt{sb} on \texttt{exampleProc} for the outcome where both loads return 0. Each column in the graph represents the path of one litmus test instruction through the microarchitecture, and each row represents the nodes corresponding to one type of event being modelled. So for instance, the node in the second row and first column of the graph represents instruction \texttt{i1} at its \texttt{Execute} stage, while the node in the second row and second column represents instruction \texttt{i2} at its \texttt{Execute} stage.

Edges in $\mu$hb graphs represent happens-before relationships between nodes, and correspond to the various orderings the microarchitecture enforces. e.g. the blue edge between the \texttt{Execute} stages of \texttt{i1} and \texttt{i2} indicates that \texttt{i1} goes through \texttt{Execute} before \texttt{i2} does, as required of an in-order pipeline.

The last two rows in the $\mu$hb graph (\texttt{L1ViCL\_C} and \texttt{L1ViCL\_E}) use the ViCL (Value in Cache Lifetime) abstraction to model cache occupancy and coherence protocol events relevant to MCM verification. Briefly speaking, a ViCL represents the period of time (relative to a single cache or main memory) over which a given cache/memory line provides a specific value for a specific address. The time period referenced by a ViCL begins at a \textit{ViCL Create} event, and ends at a \textit{ViCL Expire} event.
Figure~\ref{fig:sb_uhb} uses \texttt{L1ViCL\_C} and \texttt{L1ViCL\_E} to refer to ViCL Create and ViCL Expire events respectively, for the L1 caches in \texttt{exampleProc}. We refer the reader to CCICheck~\cite{manerkar:ccicheck} for further details.

$\mu$hb graphs can be constructed based on microarchitectural specifications that dictate when and where nodes and edges should be added for a given program. Prior work specified microarchitectural orderings as a set of axioms in the domain-specific $\mu$spec language~\cite{lustig:coatcheck}, which is similar to first-order logic. Figure~\ref{fig:example_axiom} shows an example $\mu$spec axiom. This axiom enforces that for every \texttt{microop} \texttt{i} (a \texttt{microop} is a single load or store), if it is a load (enforced through the \texttt{IsAnyRead} predicate) that reads from the initial state of memory (\texttt{DataFromInitialState i}), then its L1 ViCL must expire before the creation of L1 ViCLs of any write \texttt{j} to that address, as the write would have caused the invalidation of all other cache lines for that address. The loads \texttt{i2} and \texttt{i4} in \texttt{sb} both read from the initial state, so this axiom adds the red edges in Figure~\ref{fig:sb_uhb} to enforce the expiration of their ViCLs before the creation of ViCLs for stores \texttt{i3} and \texttt{i1} respectively. $\mu$spec supports both \texttt{forall} and \texttt{exists} quantifiers. \texttt{forall} quantifiers are translated to ANDs over the litmus test instructions, while \texttt{exists} quantifiers are translated to ORs over the litmus test instructions.

Other axioms for \texttt{exampleProc} would enforce that a given read or write goes through the pipeline stages in order, and that instructions on the same core (like \texttt{i1} and \texttt{i2}) proceed through a given pipeline stage in program order (blue edges in Figure~\ref{fig:sb_uhb}). Another axiom would enforce that a write must create its L1 ViCL (i.e. reach the memory hierarchy) before subsequent loads execute (in order to maintain program order as required by SC), shown by the yellow edges in Figure~\ref{fig:sb_uhb}.

A cycle in a $\mu$hb graph implies that an event must happen before itself, which is impossible. Thus, cyclic $\mu$hb graphs correspond to executions that are unobservable on the target microarchitecture. Likewise, acyclic $\mu$hb graphs correspond to executions that are observable on the target microarchitecture. Figure~\ref{fig:sb_uhb}'s $\mu$hb graph is cyclic (the edges that comprise the cycle are bolded), and thus this execution is unobservable on \texttt{exampleProc}, as required of an SC microarchitecture.

Typically, there are multiple $\mu$hb graphs for a given litmus test and microarchitecture. These correspond to different possible event orderings. To automatically verify a given litmus test on a microarchitecture, prior work uses SMT~\cite{barrett:handbook} solvers to search for an acyclic $\mu$hb graph that satisfies all $\mu$spec axioms. If such a graph is found, the litmus test is observable on the microarchitecture, while if no such graph can be found, the litmus test is guaranteed to be unobservable on the microarchitecture. The microarchitectural observability of the litmus test is then compared to ISA-level MCM requirements. In particular, if the observable litmus test is forbidden by the ISA-level MCM, then the microarchitecture is buggy.

\subsection{Deficiencies of Prior Automated Approaches}
\label{sec:omniscience}

\subsubsection{No Scoping of Axioms}
The flat verification conducted by prior automated microarchitectural MCM verification approaches encourages the use of axioms which exercise a global view of the entire processor.
For instance, Figure~\ref{fig:example_axiom}'s axiom adds $\mu$hb edges between ViCLs of Core 0's L1 and Core 1's L1 in Figure~\ref{fig:sb_uhb} to reflect that the loads must read memory before the stores invalidate their cache lines through the coherence protocol.
While this axiom is straightforward to write and would be valid in a coherent memory hierarchy, it does not correspond to what the hardware is actually doing, because no single hardware component has global visibility of the entire processor. For example, in Figure~\ref{fig:relcheck_pillars}b's processor, Core 0's L1 can observe the values in its own cache, but it cannot see the values in Core 1's L1. Each component in a hardware design can only enforce orderings on the events that it sees.
The ordering in Figure~\ref{fig:example_axiom}'s axiom is actually enforced in a distributed manner by a combination of modules, specifically the L1s and (likely) a bus or directory.
An architect may surmise that the combination of the orderings enforced by the L1s and bus/directory is sufficient to enforce Figure~\ref{fig:example_axiom}'s axiom, but such an assumption must be validated before using the axiom for microarchitectural MCM verification. Otherwise it is possible that the axiom does not hold in the actual hardware design, leading to unsound verification.
A better approach is to allow the teams working on each module to specify the axioms that actually hold in their modules, and then verify that the composition of the modules correctly maintains required MCM orderings. However, all axioms in prior work have global visibility. There is no way for users to write axioms that are scoped to one portion of the design.

\subsubsection{No Per-Module or Scalable Verification}
Prior automated MCM verification approaches also provided no way for users to verify a component independently of the rest of the system. For instance, all verification in Figure~\ref{fig:sb_uhb}'s $\mu$hb graph is conducted in terms of instructions, relative to an ISA-level litmus test. There is no way to verify say, the L1 caches individually.
Since prior work can only verify a design all at once, it cannot scale to large detailed designs due to the exponential complexity of SMT solving~\cite{barrett:handbook}.
Ideally, each component would have an interface specification that it could be verified against, and it would be possible to verify a design piece-by-piece, allowing verification to scale to large detailed designs.

RealityCheck solves the above problems through its approach to modularity, hierarchy, and abstraction. The next section provides a high-level overview of how it does so.


\section{RealityCheck Overview}
\label{sec:realitycheck}

\begin{figure*}[t]
  \centering
  \includegraphics[width=0.8\textwidth]{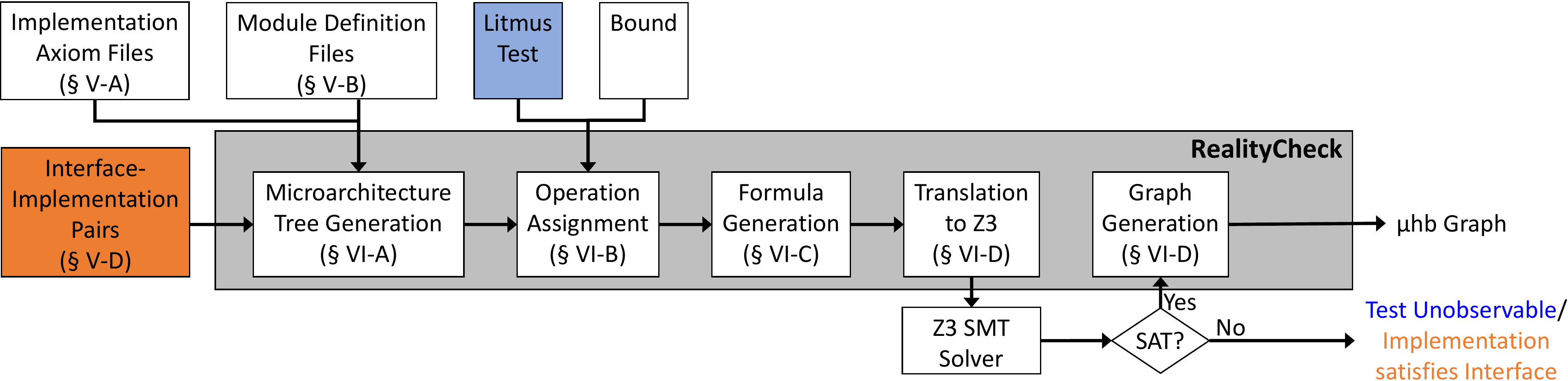}
  \vspace{-10pt}
  \caption{RealityCheck block diagram. Orange parts only apply to interface verification, and blue parts only to litmus test verification.}
  \label{fig:realitycheck}
  \vspace{-10pt}
\end{figure*}

\begin{figure}[t]
  \centering
  \includegraphics[width=0.45\textwidth]{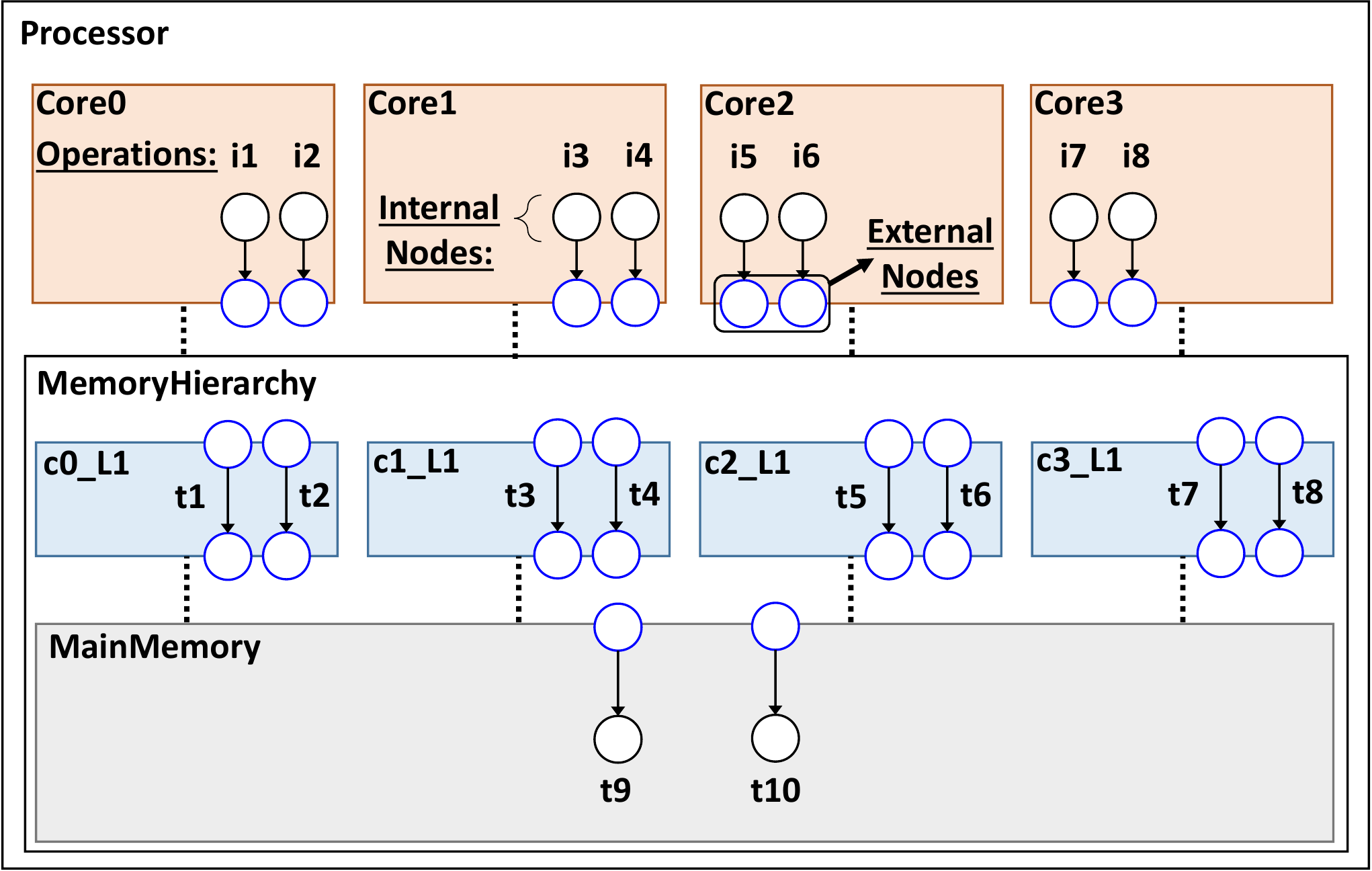}
  \vspace{-5pt}
  \caption{High-level graphical depiction of RealityCheck's model of Figure~\ref{fig:relcheck_pillars}c's processor, showing examples of operations, internal nodes, and external nodes. \texttt{MemoryHierarchy}'s operations are not shown for brevity.}
  \label{fig:relcheck_basics}
  \vspace{-5pt}
\end{figure}

\begin{figure}[t]
  \centering
  \includegraphics[width=0.49\textwidth]{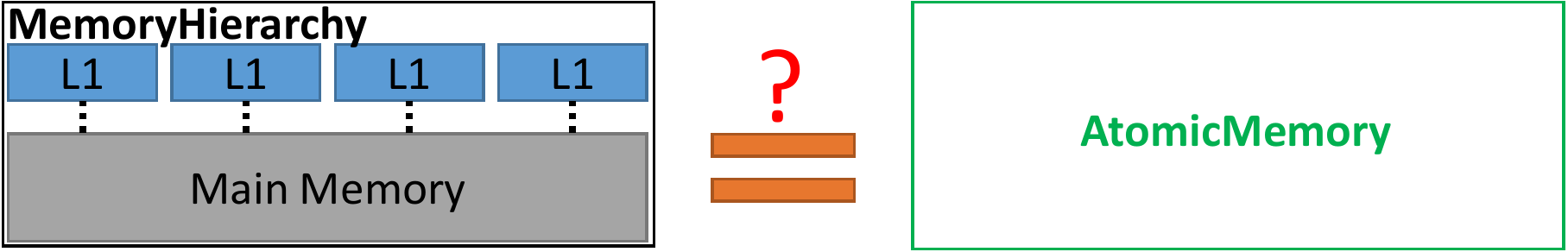}
  \vspace{-15pt}
  \caption{Interface verification: checking a set of implementation modules against an abstract interface.}
  \label{fig:interface_verif}
  \vspace{-10pt}
\end{figure}

Figure~\ref{fig:realitycheck} shows the high-level block diagram of RealityCheck. RealityCheck can be run in one of two ways: (i) for litmus test verification to verify a modular microarchitectural ordering specification against an ISA-level litmus test, or (ii) for interface verification to verify the microarchitectural ordering specification of design components against the ordering specification of their abstract interfaces (Section~\ref{sec:interface_benefits}). The latter use case enables a module to be verified independently of the rest of the design.
The five steps in RealityCheck operation (\textbf{Microarchitecture Tree Generation}, \textbf{Operation Assignment}, \textbf{Formula Generation}, \textbf{Translation to Z3}, and \textbf{Graph Generation}) are common to both litmus test verification and interface verification. The difference between the two cases lies in which modules are checked and which operations they are checked on.
Operations represent the instructions or instruction-like quantities that a module operates on. For example, the operations of a core module would be instructions, but the operations of a memory module would be memory transactions. Figure~\ref{fig:relcheck_basics} provides a high-level graphical depiction of RealityCheck's basic terms (including operations) for Figure~\ref{fig:relcheck_pillars}c's processor.

Two inputs that are provided to RealityCheck in both litmus test verification and interface verification are the implementation axiom files and the module definition files. These files are specified in the $\mu$spec++ language (Section~\ref{sec:uspecplusplus}) developed as part of RealityCheck. The $\mu$spec++ language is based on the $\mu$spec language~\cite{lustig:coatcheck}, but adds support to the language for modularity, hierarchy, and abstraction, much like C++ does to C.
The module definition files (Section~\ref{sec:uarch_defn}) specify $\mu$spec++ modules in a manner similar to a C++ \texttt{.h} file.
Meanwhile, each implementation axiom file (Section~\ref{sec:impl_axioms}) specifies the events relevant to a given module as well as orderings on these events, in a manner similar to a C++ \texttt{.cpp} file.

If verifying a litmus test, the test is provided as input in the \texttt{.test} format from prior work~\cite{lustig:coatcheck}. Meanwhile, if running interface verification, RealityCheck takes in a list of implementation-interface pairs. Each pair specifies an implementation module to verify against an interface specification, and their corresponding node mappings (Section~\ref{sec:interface_spec}).

The final input to RealityCheck (which is always provided to the tool) is the bound up to which to conduct verification. Similar to most prior automated hardware MCM verification work~\cite{lustig:pipecheck,manerkar:ccicheck,lustig:coatcheck,trippel:tricheck,manerkar:rtlcheck} but unlike PipeProof~\cite{manerkar:pipeproof}, RealityCheck conducts bounded verification; i.e. it explores all possible executions that use up to the specified number of operations (per module).
Thus, RealityCheck is excellent for bug-finding, as we show in our case studies in Section~\ref{sec:bugfinding}.




\section{Abstraction and Its Benefits}
\label{sec:interface_benefits}

In RealityCheck, interfaces can be used to separate the specification of a component's functional behaviour from the details of its implementation. For example, users may want to abstract the behaviour of their memory hierarchy as a single atomic memory, as shown in Figure~\ref{fig:interface_verif}.

The use of interfaces has several benefits.
First, any implementation of the interface can be verified against the interface specification independently of the rest of the system. This gives users a method to verify the correctness of a design component without needing to link it to a top-level litmus test. For example, Figure~\ref{fig:interface_verif} shows the verification of the \texttt{MemoryHierarchy} module and its submodules (instances of other modules that exist within \texttt{MemoryHierarchy}) against the \texttt{AtomicMemory} interface. Interface verification enables easy localisation of bugs to a given module based on whether it satisfies its interface.
Second, interfaces enable scalable verification. Instead of verifying the entire design at once (Figure~\ref{fig:relcheck_pillars}c), which will likely result in an SMT formula too large for solvers to handle, interfaces enable verification to be split into multiple steps. Specifically, the design is first verified using the (likely) smaller and simpler interface specification of the component (Figure~\ref{fig:relcheck_pillars}d) rather than its implementation. Then, the component implementation is separately verified against the interface specification up to a user-provided bound (Figure~\ref{fig:interface_verif}). These two verification queries can be run in parallel, and will likely be smaller SMT formulae than verifying the design all at once. This process can be repeated further down the hierarchy. For instance, if the L1 caches in Figure~\ref{fig:interface_verif} had interfaces, those interfaces could be used when conducting interface verification in Figure~\ref{fig:interface_verif}. The L1 implementation could then be separately verified against its interface specification. This splitting of verification queries using interfaces can be done again and again to split verification into smaller problems, thus allowing it to scale.

Third, interfaces allow implementations to be switched in and out easily. For example, if the user wants to introduce a new memory hierarchy (say, one with an L2) to a previously verified version of Figure~\ref{fig:relcheck_pillars}d, then all they need to do to ensure correctness is to verify the new memory hierarchy against the \texttt{AtomicMemory} interface, independently of the rest of the design.
Finally, the use of interfaces facilitates sharing of IP between vendors at the SoC level. A vendor can internally verify their implementation against its interface for correctness, and then share their interface with other vendors without having to share their internal implementation specification.

\section{$\mu$spec++ Modular Design Specifications}
\label{sec:uspecplusplus}

This section explains the $\mu$spec++ domain-specific language using a pedagogical microarchitecture \texttt{simpleProc}, comprised of 3-stage pipelines connected to a single main memory.

\subsection{Implementation Axiom Files}
\label{sec:impl_axioms}

\begin{figure}[t]
  \centering
  \begin{small}
  \begin{verbatim}
ModuleID "Core".

DefineEvent 0 "IF".
DefineEvent 1 "EX".
DefineEvent 2 "WB".
DefineEvent External 3 "MemReq".
DefineEvent External 4 "MemResp".

Axiom "PO_Fetch":
forall microop "i1",
forall microop "i2",
ProgramOrder i1 i2 =>
  AddEdge ((i1, IF), (i2, IF), "").

Axiom "Req_Resp_PO":
forall microop "i1",
forall microop "i2",
EdgeExists ((i1, WB), (i2, WB), "") /\
NodesExist [(i1, MemResp); (i2, MemReq)] =>
  AddEdge ((i1, MemResp), (i2, MemReq),"").
\end{verbatim}
\end{small}
  \vspace{-10pt}
  \caption{Part of \texttt{simpleProc}'s \texttt{Core} module's implementation axiom file.}
  \vspace{-10pt}
  \label{fig:impl_axioms}
\end{figure}

Each implementation axiom file specifies the events relevant to a given module. Figure~\ref{fig:impl_axioms} shows part of the implementation axiom file for a module of type \texttt{Core}. The file begins with the module's type, and is followed by a list of the types of events that this module can observe and/or enforce orderings on, denoted using \texttt{DefineEvent}. The rest of the file details the axioms which enforce orderings on these events. Figure~\ref{fig:impl_axioms} shows two such axioms, \texttt{PO\_Fetch} and \texttt{Req\_Resp\_PO}. These axioms are identical to $\mu$spec axioms, except that their scope is restricted to the operations of the module in which they are declared. For example, if evaluating Figure~\ref{fig:sb_code}'s litmus test on a $\mu$spec++ design, an instance of the \texttt{Core} module representing Core 0 would only be able to see instructions \texttt{i1} and \texttt{i2} rather than all the instructions of the test. In such a case, the \texttt{forall} quantifiers in the \texttt{PO\_Fetch} axiom would evaluate to an AND over instructions \texttt{i1} and \texttt{i2}.

By default, events can only be viewed by the module instance in which they are declared, similar to private member variables in C++. For example, the \texttt{IF} and \texttt{WB} events in an instance of the \texttt{Core} module cannot be seen outside that \texttt{Core} instance.
This capability allows designers to hide events internal to their design component from other modules in the system, just like in a real Verilog design.

Meanwhile, a module's external events can be viewed by itself as well as by its parent module and any modules it may be connected to (see Section~\ref{sec:uarch_defn} for further details). Such events are labelled with the \texttt{External} keyword when they are declared in the implementation axiom file. For example, the \texttt{MemReq} and \texttt{MemResp} events in Figure~\ref{fig:impl_axioms}'s \texttt{Core} module are both external events. Thus, if the \texttt{Req\_Resp\_PO} axiom adds an edge between the \texttt{MemResp} and \texttt{MemReq} nodes of two instructions, that edge will be visible outside the instance of the \texttt{Core} module in which the instructions reside. Figure~\ref{fig:relcheck_basics} shows a graphical depiction of internal and external events/nodes.

\subsection{Module Definition Files}
\label{sec:uarch_defn}

\begin{figure}[t]
  \centering
  \begin{small}
  \begin{verbatim}
Module Processor () {
  OperationType none
  Properties  { IsCore no }
  
  Submodules  {
    Core c0 (c : 0)
    Core c1 (c : 1)
    Core c2 (c : 2)
    Core c3 (c : 3)
    Mem mem ()
  }

  ConnectionAxioms {
    Axiom "instr_has_tran":
    forall microop "i" in "c0;c1;c2;c3",
    NodeExists (i, MemReq) =>
      exists transaction "j" in "mem",
        Mapped i j.
    ...
    Axiom "mapped_effects":
    forall microop "i" in "c0;c1;c2;c3",
    forall transaction "j" in "mem",
    Mapped i j =>
     (SameAddress i j /\ SameData i j /\
      (IsAnyRead i <=> IsAnyRead j) /\
      (IsAnyWrite i <=> IsAnyWrite j) /\
      SameNode (i, MemReq) (j, Req) /\
      SameNode (i, MemResp) (j, Resp)).
  }}
\end{verbatim}
\end{small}
  \vspace{-10pt}
  \caption{\texttt{simpleProc}'s \texttt{Processor} module definition.}
  \vspace{-10pt}
  \label{fig:uarch_defn_eg}
\end{figure}

Figure~\ref{fig:uarch_defn_eg} shows the module definition of \texttt{simpleProc}'s top-level \texttt{Processor} module.
A module definition file specifies the module's operation type, properties, submodules, and connection axioms.

\subsubsection{Operation Types and Properties}

The operations in each module have some type, specified using the \texttt{OperationType} keyword. For example, \texttt{simpleProc}'s \texttt{Core} module has an operation type of \texttt{microop}, since it deals with instructions, while the \texttt{Mem} module has an operation type of \texttt{transaction}. Users can add additional operation types.
The \texttt{Processor} module in Figure~\ref{fig:uarch_defn_eg} has an operation type of \texttt{none}, a special identifier indicating that it has no operations. This is because the \texttt{Processor} module serves only to encapsulate the other modules in the system.

When quantifying over operations, RealityCheck verifies that the operation type used in the quantifier matches that of the operations over which the quantifier is being evaluated. So for instance, if the first \texttt{forall} quantifier in Figure~\ref{fig:impl_axioms}'s \texttt{PO\_Fetch} axiom was replaced with \texttt{forall transaction "i1"}, RealityCheck would flag a type error.

A module's properties are certain fields that are shared across all instances of the module, similar to static variables of classes in C++. An example of a property is the \texttt{IsCore} property, which can be set to \texttt{yes} or \texttt{no}.

\subsubsection{Submodules}
\label{sec:submodules}

\begin{figure}[t]
  \centering
  \begin{small}
  \begin{verbatim}
Axiom "Mem_Writes_Path":
forall transaction "i",
    NodeExists (i, Req) /\ IsAnyWrite i =>
    AddEdges [((i, Req), (i, ViCL_C), "");
              ((i, ViCL_C), (i, ViCL_E), "");
              ((i, ViCL_C), (i, Resp), "")].
\end{verbatim}
\end{small}
  \vspace{-10pt}
  \caption{An implementation axiom of \texttt{simpleProc}'s \texttt{Mem}.}
  \vspace{-15pt}
  \label{fig:mem_impl_axiom}
\end{figure}

The submodules of a module are instances of other modules that exist within it. Submodules enable hierarchy in RealityCheck, allowing larger modules to be built using smaller ones. A module can evaluate $\mu$spec++ predicates on the operations of its submodules and observe their external events, but it cannot observe their internal events.

When instantiating a submodule, parameters may be passed to the instance to populate some instance-specific fields (similar to C++ constructors). For example, when the \texttt{Processor} module instantiates \texttt{Core} modules as its submodules, it passes each of them an integer parameter \texttt{c} denoting their core ID. This parameter can then be used in the axioms of that module.


\subsubsection{Connection Axioms}
\label{sec:conn_axioms}

\begin{figure}[t]
  \centering
  \includegraphics[width=0.4\textwidth]{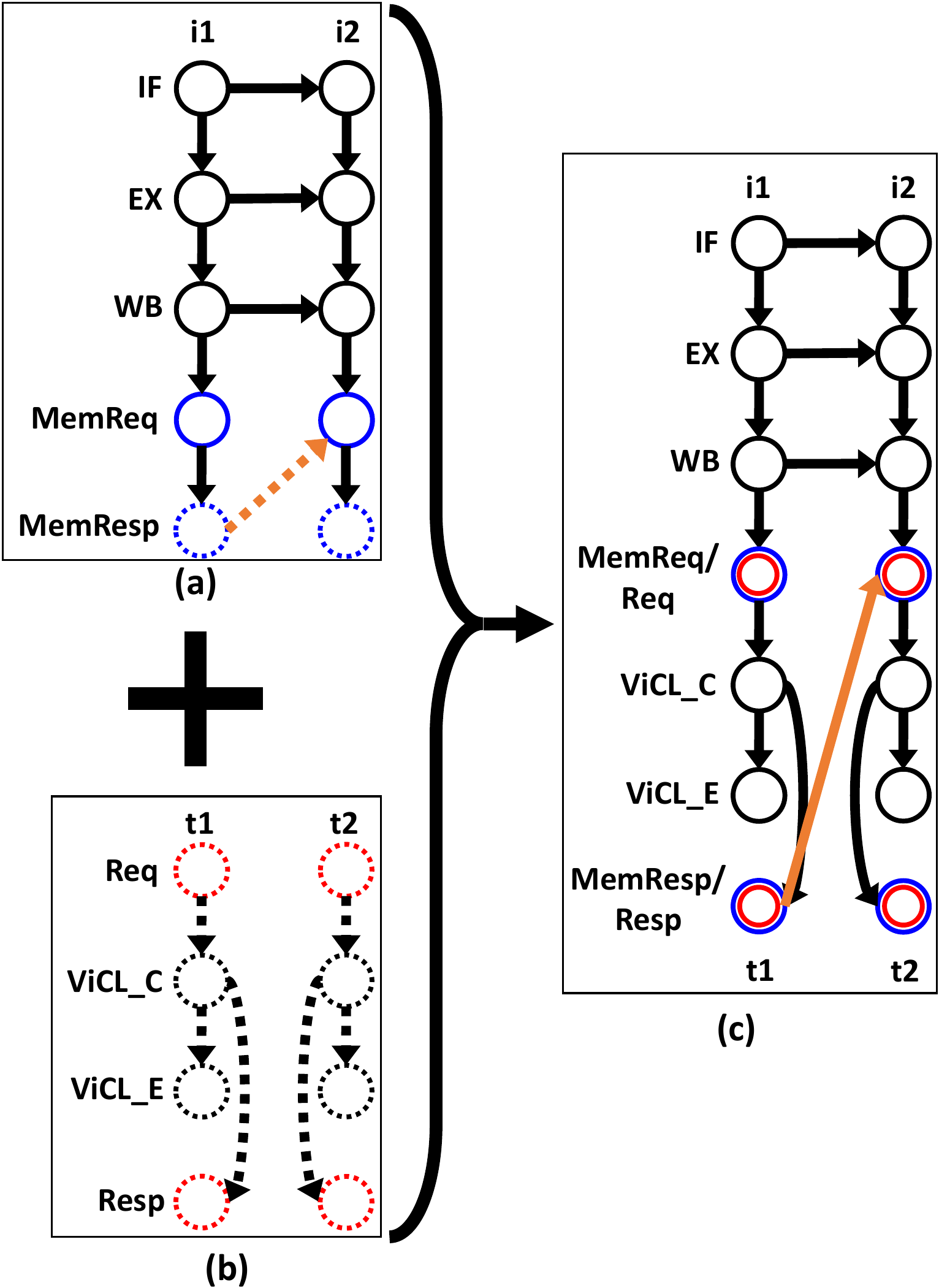}
  \caption{Effect of connection axioms in \texttt{simpleProc}, assuming 3-stage pipelines. (a) shows two stores in program order from a \texttt{Core}, while (b) shows two transactions from \texttt{Mem}. (c) shows the result when connection axioms merge the \texttt{MemReq} and \texttt{MemResp} nodes of \texttt{Core} with the \texttt{Req} and \texttt{Resp} nodes of \texttt{Mem}. Black nodes are internal nodes while red and blue nodes are external nodes. Dotted nodes and edges are not guaranteed to exist. Merged nodes are concentric circles.}
  \label{fig:samenode}
  \vspace{-5pt}
\end{figure}


Submodules are connected to each other and to their parent module through connection axioms.
The bottom of Figure~\ref{fig:uarch_defn_eg} shows two example connection axioms. Connection axioms are similar to implementation axioms (Section~\ref{sec:impl_axioms}), but have differences in their scope. A module's connection axioms can observe the operations of the module itself and those of its submodules. They can observe all events of their module (both internal and external), but only the external events of any submodules.

Since connection axioms can observe the operations of multiple modules, their quantifiers must specify the domain over which they operate. Each quantifier provides a list of modules whose operations it applies to (\texttt{this} refers to the operations of the module itself). For example, the \texttt{forall} quantifier in the \texttt{instr\_has\_tran} axiom is evaluated on the operations from modules \texttt{c0}, \texttt{c1}, \texttt{c2}, and \texttt{c3}. Thus, the quantifier evaluates to an AND over all these operations, but does not apply to the operations in the \texttt{mem} module.

Connection axioms are responsible for translating one module's operations to those of another, and for linking them together. For example, if a core is connected to memory, the core's instructions need to be translated and mapped to their corresponding memory transactions. Furthermore, there may be multiple possible translations for a given litmus test. e.g. a load may read from the store buffer in one execution (and thus generate no memory transactions), while in another execution it may read from memory, thus generating a memory transaction.

By checking all the different ways the connection axioms could be satisfied, RealityCheck examines all possible translations of operations between modules.
For example, the \texttt{instr\_has\_tran} axiom in Figure~\ref{fig:uarch_defn_eg} maps each instruction on the four cores which requests data from memory (signified by the existence of its external \texttt{MemReq} node) to some transaction in \texttt{mem}, denoted by the \texttt{Mapped} predicate. This reflects how in a real design, an instruction in the pipeline that accesses memory will generate a corresponding memory transaction. Other axioms not shown ensure that the mapping between instructions and memory transactions is 1-1. Mapping schemes other than 1-1 can be used where necessary.

If operations are mapped to each other 1-1, mapped pairs must agree on which addresses, values, etc are being accessed. They must also agree on the timing of their events. For example, the second connection axiom (\texttt{mapped\_effects}) in Figure~\ref{fig:uarch_defn_eg} enforces some of these orderings. It enforces that if an instruction is mapped to a memory transaction, then they must have the same address and read/write the same data value. It also enforces that load instructions map to read transactions and stores map to write transactions (through the \texttt{IsAnyRead} and \texttt{IsAnyWrite} predicates). In addition, it uses the \texttt{SameNode} predicate to link the \texttt{MemReq} event of the instruction \texttt{i} to the \texttt{Req} event of the transaction \texttt{j}. Likewise, the \texttt{MemResp} event of \texttt{i} is linked to the \texttt{Resp} event of \texttt{j}.

Linking two nodes with \texttt{SameNode} essentially merges the two nodes together, ensuring that they are exactly the same event. In this case, the processor's request to and response from memory are viewed from the memory side as an arriving request to which it sends a response. Further details about the semantics of \texttt{SameNode} are discussed in Section~\ref{sec:z3}.

\subsubsection{A Graphical Example}

Figure~\ref{fig:samenode} graphically depicts the effect of connection axioms in \texttt{simpleProc}. Black outlines denote internal nodes. External nodes are outlined in blue (in \texttt{Core}) or red (in \texttt{Mem}). In Figure~\ref{fig:samenode}a we have two stores \texttt{i1} and \texttt{i2} in program order from an instance of \texttt{Core}. Note that the \texttt{MemResp} nodes of \texttt{i1} and \texttt{i2} are dotted to indicate that these nodes are not guaranteed to exist. This reflects the fact that a \texttt{Core} cannot just assume that its memory requests will be responded to. The existence of the \texttt{MemResp} nodes must be enforced by axioms in the \texttt{Mem} module and communicated to the \texttt{Core} through connection axioms. The implementation axioms of the \texttt{Core} module in Figure~\ref{fig:impl_axioms} obey this convention; for instance, the \texttt{Req\_Resp\_PO} axiom does not add an edge between the \texttt{MemResp} event of \texttt{i1} and the \texttt{MemReq} event of \texttt{i2} unless the \texttt{MemResp} node of \texttt{i1} exists.

Meanwhile, in Figure~\ref{fig:samenode}b we have two memory transactions \texttt{t1} and \texttt{t2}, governed by the axioms of the \texttt{Mem} module -- specifically, the axiom shown in Figure~\ref{fig:mem_impl_axiom}. This axiom enforces that if a write request is provided to the \texttt{Mem} module, then it is responded to. Note that all the nodes of \texttt{t1} and \texttt{t2} are dotted, indicating that none of them are guaranteed to exist. This reflects the fact that in \texttt{simpleProc}, memory will remain idle unless data is provided to or requested from it. Without connection axioms, no instructions must interact with memory, and so no nodes or edges in \texttt{Mem} are guaranteed to exist.

When the connection axioms in Figure~\ref{fig:uarch_defn_eg} are enforced, the result is Figure~\ref{fig:samenode}c, where instruction \texttt{i1} is mapped to transaction \texttt{t1} and instruction \texttt{i2} is mapped to transaction \texttt{t2}. The \texttt{Req} node of transaction \texttt{t1} is now guaranteed to exist, because it is the same node as \texttt{i1}'s \texttt{MemReq} node (denoted by the concentric blue and red circles), which is guaranteed to exist by \texttt{Core}. Transaction \texttt{t2}'s \texttt{Req} node is similarly guaranteed to exist by \texttt{t2} being mapped to \texttt{i2}. Figure~\ref{fig:mem_impl_axiom}'s axiom now enforces that both transactions are responded to, causing the \texttt{Resp} nodes of \texttt{t1} and \texttt{t2} to exist, and thus also causing the \texttt{MemResp} nodes of \texttt{i1} and \texttt{i2} to exist (since they are now merged with the \texttt{Resp} nodes). Finally, since the \texttt{MemResp} node of \texttt{i1} now exists, the \texttt{Req\_Resp\_PO} axiom of the \texttt{Core} module (Figure~\ref{fig:impl_axioms}) now enforces (through the orange edge in Figure~\ref{fig:samenode}) that the \texttt{Core} must receive \texttt{i1}'s response from memory before sending \texttt{i2}'s request to memory.

\subsection{Preventing Globally Scoped Axioms}
\label{sec:prev_omniscience}

Consider trying to include the globally scoped axiom from Figure~\ref{fig:example_axiom} in a RealityCheck specification for Figure~\ref{fig:relcheck_pillars}c's processor. Figure~\ref{fig:example_axiom}'s axiom references L1 ViCL nodes, which are internal to their corresponding L1 instance in the processor's module hierarchy. Thus, an axiom that references ViCL nodes (like Figure~\ref{fig:example_axiom}) must be an internal axiom in the L1 module. However, any internal axiom in an L1 module instance can only observe its own transactions.
For example, if evaluating the litmus test \texttt{sb} (Figure~\ref{fig:sb_code}), the internal axioms of Core 0's L1 module can only see the memory transactions mapped to \texttt{i1} and \texttt{i2}.
$\mu$spec++ provides no way for Core 0's L1's internal axioms to refer to the memory transactions of \texttt{i3} and \texttt{i4}. Similarly, Core 1's L1 instance can observe the memory transactions of \texttt{i3} and \texttt{i4}, but not those of \texttt{i1} and \texttt{i2}.
Thus, as an internal axiom in RealityCheck, Figure~\ref{fig:example_axiom} cannot enforce Figure~\ref{fig:sb_uhb}'s red edges between \texttt{i2} and \texttt{i3} or between \texttt{i4} and \texttt{i1}, because it cannot refer to either of these pairs at the same time. Likewise, the internal or connection axioms of the \texttt{MemoryHierarchy} module can refer to the memory transactions of all litmus test instructions, but they cannot refer to the internal ViCL nodes of the L1 modules (RealityCheck will flag an error if they do so). Either way, RealityCheck makes it impossible to write a single axiom that accomplishes Figure~\ref{fig:example_axiom}'s functionality. The ordering must be enforced by a combination of axioms: one or more connection axioms to communicate coherence invalidations from one L1 to another, and another internal axiom in the L1 module to process these invalidations and ensure that the ViCLs expire appropriately.
Thus, if users organise their $\mu$spec++ specifications into modules that reflect the structure of their design, RealityCheck will automatically prevent them from writing globally scoped axioms that violate this design structure.



\subsection{Interface Specification and Node Mappings}
\label{sec:interface_spec}

Interfaces are specified in RealityCheck in a manner similar to other modules, but with some additional constraints.
Interfaces cannot have submodules or connection axioms, as their goal is to provide a simple specification of component behaviour that does not delve into implementation details.

When verifying an implementation against an interface, the event types of the implementation must be mapped to those of the interface, so that the interface's properties can be checked on the implementation. Otherwise the interface and implementation would be referring to different events. For example, if verifying \texttt{MemoryHierarchy} against \texttt{AtomicMemory} as per Figure~\ref{fig:interface_verif}, one might map the request and response events of the memory hierarchy to the corresponding request and reponse events of the atomic memory. This list of node mappings must be provided along with a interface-implementation pair when it is input to RealityCheck for interface verification.
\section{RealityCheck Operation}
\label{sec:realitycheckop}



\subsection{Step 1: Microarchitecture Tree Generation}
\label{sec:uarch_tree_gen}

\textbf{Microarchitecture Tree Generation} creates a tree of $\mu$spec++ module instances (i.e. copies) according to the module definition files and interface/implementation axiom files. 



\subsection{Step 2: Operation Assignment}
\label{sec:op_assn}

\textbf{Operation Assignment} generates and assigns operations to each module. The design's axioms are subsequently evaluated over these operations, with the visibility restrictions enforced by $\mu$spec++ detailed earlier.

RealityCheck assigns a number of operations to each module equal to the bound specified by the user as input. For instance, if assigning operations to Figure~\ref{fig:relcheck_pillars}b's processor for a bound of 4, there would be 4 operations assigned to each of the 4 cores, each of the L1s, and to main memory, for a total of 16 + 16 + 4 = 36 operations for the entire design. The bound is the maximum number of operations per module that can exist in any verified execution, so an execution may use only some of the operations per module. RealityCheck accomplishes this by associating every operation with an implicit \texttt{IsNotNull} predicate, and enforcing that axioms only apply to non-null operations. This is the approach used by tools like Alloy~\cite{jackson:alloy}.

In litmus test verification, the design's \texttt{Core} modules (identified by the \texttt{IsCore} property) are assigned the litmus test instructions corresponding to their core. These litmus test instructions are \textit{concrete}; their type, address, and value are dictated by the litmus test and cannot change.
However, as Section~\ref{sec:conn_axioms} covers, verification must cover all possible translations (up to a bound) of these litmus test instructions to lower-level modules.
RealityCheck enables such translation by having operations in modules other than cores be \textit{symbolic}, and having connection axioms enforce requirements on them based on the instructions they are (directly or indirectly) mapped to. Symbolic operations are abstract operations which can have any type (e.g.: read, write, etc.), address, or value, as long as the design's axioms are maintained.


Meanwhile, in interface verification, all operations of involved modules are symbolic. Thus, in that case, RealityCheck verifies that an implementation satisfies its interface for all possible combinations of operations up to the bound.

\subsection{Step 3: Formula Generation}
\label{sec:form_gen}

In Formula Generation,
RealityCheck takes the conjunction of every module's implementation axioms and connection axioms, eliminating quantifiers by translating \texttt{forall}s into ANDs and \texttt{exists} into ORs over each quantifier's domain (the operations being quantified over).
RealityCheck conducts some preliminary simplification on the resultant representation, and then converts it (as described below) into an SMT formula checkable by the Z3 SMT solver~\cite{demoura:z3}.

\subsection{Steps 4 \& 5: Translate to Z3 and Graph Generation}
\label{sec:z3}

RealityCheck translates AND, OR, and NOT operators to their Z3 equivalents. Each predicate is mapped to a Z3 Boolean variable, except for \texttt{SameNode} (explained below).
RealityCheck uses Z3's Linear Integer Arithmetic (LIA) theory to enforce happens-before orders. Each $\mu$hb node has two variables in Z3. The first is a Boolean variable dictating whether or not the node exists. The second is an integer variable denoting the timestamp of the node in the microarchitectural execution. An edge from a node \texttt{s} to a node \texttt{d} is translated to a constraint \texttt{e\_s} $<$ \texttt{e\_d} where \texttt{e\_s} and \texttt{e\_d} are the integer variables denoting the timestamps of \texttt{s} and \texttt{d} respectively.

The \texttt{SameNode} predicate requires special handling, as it is not just a Boolean predicate, but also enforces that two nodes be merged together. If the user declares two nodes to be the same node in their $\mu$spec++, RealityCheck first creates a bi-implication between their Boolean variables to ensure that if one exists, so does the other (and vice versa). Then, RealityCheck adds a constraint that the integer variables denoting the timestamps of the nodes must have the same value. Together, these two constraints ensure that the two nodes in the \texttt{SameNode} predicate are treated as the same node.

If Z3 finds a satisfying assignment to the generated formula, the assignment represents an acyclic graph where nodes with their Boolean variables set to true exist, and where edges exist between nodes \texttt{s} and \texttt{d} if the integer variable for \texttt{s} is less than the integer variable for \texttt{d}. In such a case, RealityCheck parses the satisyfing assignment to generate a $\mu$hb graph that the user can view and use for debugging. If Z3 cannot find a satisfying assignment, then no acyclic $\mu$hb graph satisfying the constraints exists, and the outcome under consideration is unobservable (up to the user-specified bound).

\section{RealityCheck Usage Flows}
\label{sec:best_practices}

RealityCheck may be used to verify a component against its interface in isolation (Section~\ref{sec:interface_benefits}), independent of the rest of the design. This capability of per-module verification enables RealityCheck to adapt to multiple design flows. If used at early-stage design time, users may first come up with a shallow design specification where all modules are represented by their interfaces, and then progressively replace modules with their submodules and implementations to create a more detailed design over time. Interface verification can be used to check implementations for correctness as the design becomes more detailed in this ``outside-in'' approach. If interface verification finds bugs, then additional axioms should be added to the implementation until interface verification succeeds.

On the other hand, if an implementation (or RTL) already exists, users may favour an ``inside-out'' approach, where they first model one or a few modules deep in the system, and progressively add more modules and hierarchy to the specification until the entire design is modelled. In the case where RTL exists, users may use a tool like RTLCheck~\cite{manerkar:rtlcheck} to check that the axioms they are writing for modules are sound.


When decomposing a module into submodules, users should try and minimise communication between the submodules. If such decomposition proves difficult, or would result in submodules with minimal internal functionality and large amounts of communication between them, it may be better to leave the module as a single unit (i.e. no submodules). The more internal events one can hide using abstraction, the faster verification will be (generally speaking). So for instance, RealityCheck will be fast for a microarchitecture that has a simple request-response interface to memory where all other memory functionality is hidden. Likewise, it will be slower for a speculative microarchitecture where coherence invalidations are propagated up to the pipeline. Nevertheless, RealityCheck is still capable of verifying such microarchitectures.




\section{Methodology and Results}
\label{sec:results}

\begin{table}[t]
\caption{Microarchitectures Evaluated Using RealityCheck}
\centering
    \begin{tabularx}{0.49\textwidth}{ |X|X|X|p{0.05\textwidth}| } 
     \hline
     \textbf{Name} & \textbf{Pipelines} & \textbf{Mem. Hierarchy} & \textbf{MCM} \\\hline
     simpleProc & inOrderCore & unifiedMem & SC \\\hline
     cacheProc & inOrderCore & L1Hier & SC \\\hline
     simpleProcTSO & sbCore & unifiedMem & TSO \\\hline
     cacheProcTSO & sbCore & L1Hier & TSO \\\hline
     simpleProcRISCV & rvwmoCore & unifiedMem & RVWMO \\\hline
     cacheProcRISCV & rvwmoCore & L1Hier & RVWMO \\\hline
     heteroProcRISCV & 2 sbCore-RISCV, 2 rvwmoCore & L1Hier & RVWMO \\\hline
    \end{tabularx}
  \label{tab:uarches}
\end{table}

\begin{figure}[t]
  \centering
  \includegraphics[width=0.49\textwidth]{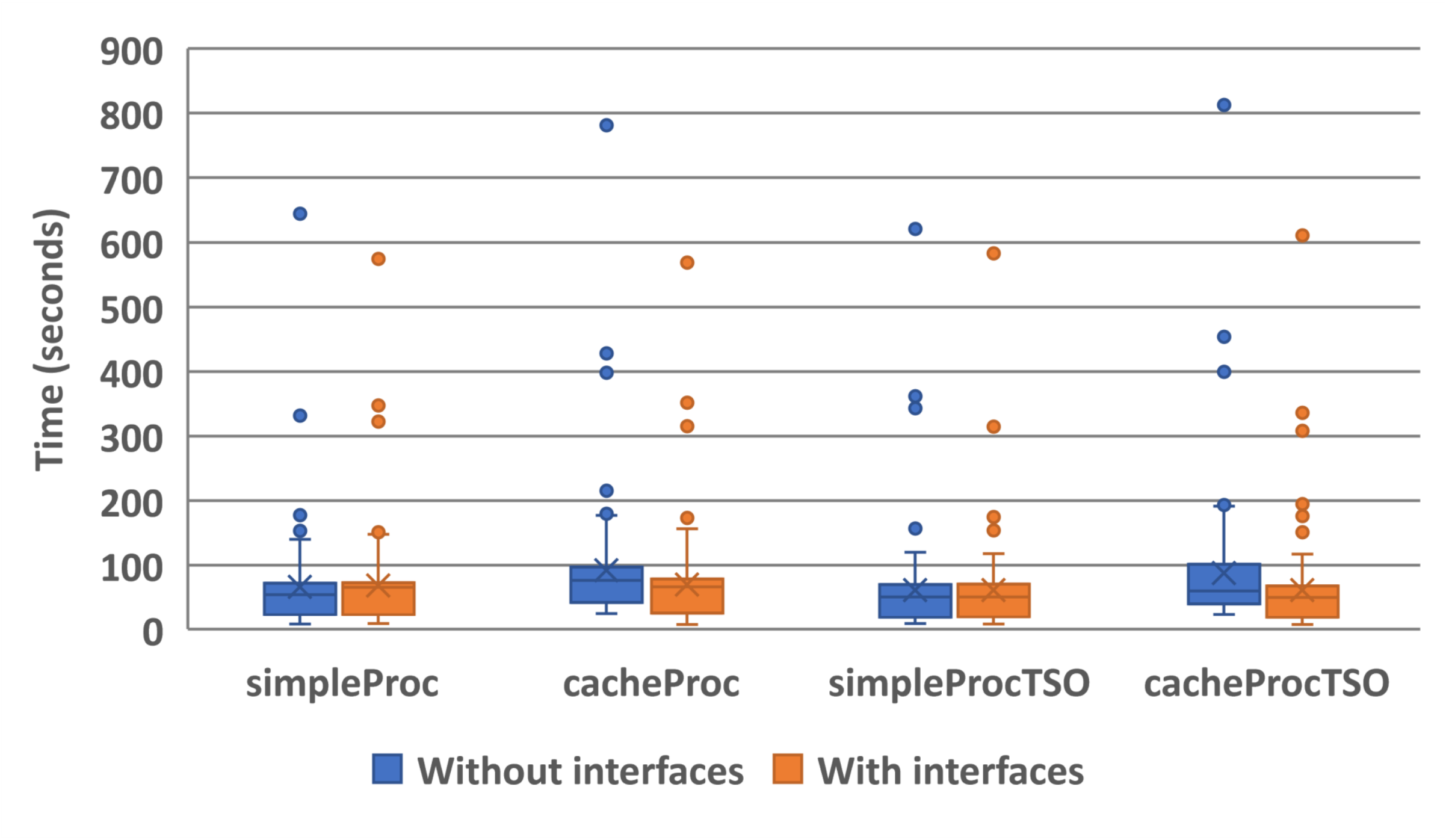}
  \vspace{-20pt}
  \caption{Verification times for 95 litmus tests across four microarchitectures implementing the SC and TSO MCMs. Each column represents the runtimes for the 95 tests for one particular configuration. The box represents the upper and lower quartile range of the data. Dots represent points lying beyond 1.5x the interquartile range (the extent of the whiskers) from the ends of the box.}
  \label{fig:runtimes_sc_tso}
  \vspace{-10pt}
\end{figure}

\begin{figure}[t]
  \centering
  \includegraphics[width=0.49\textwidth]{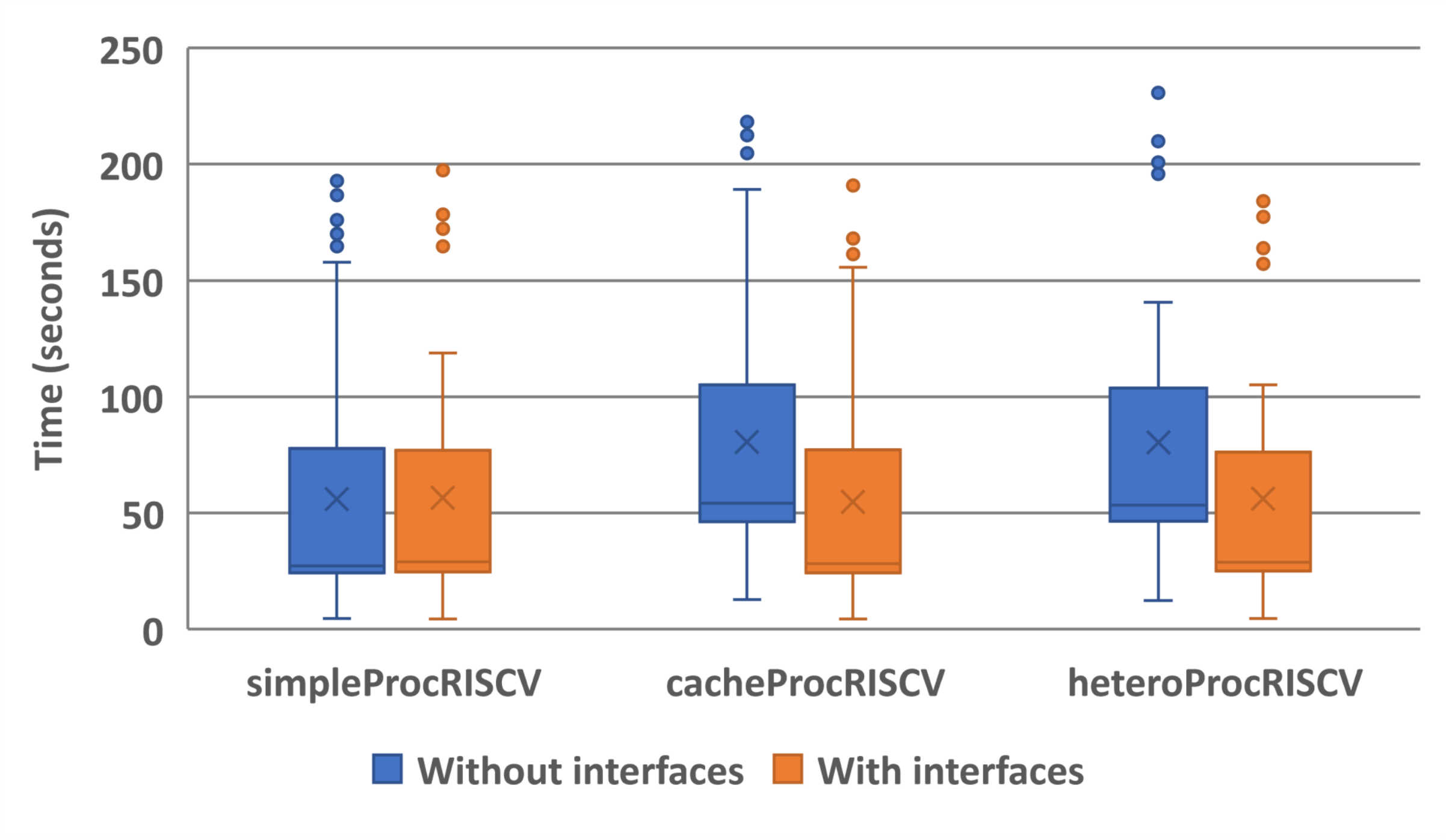}
  \vspace{-25pt}
  \caption{Verification times for 98 litmus tests across three microarchitectures implementing the RISC-V RVWMO MCM.}
  \label{fig:runtimes_riscv}
  \vspace{-5pt}
\end{figure}

\begin{figure}[t]
  \centering
  \includegraphics[width=0.35\textwidth]{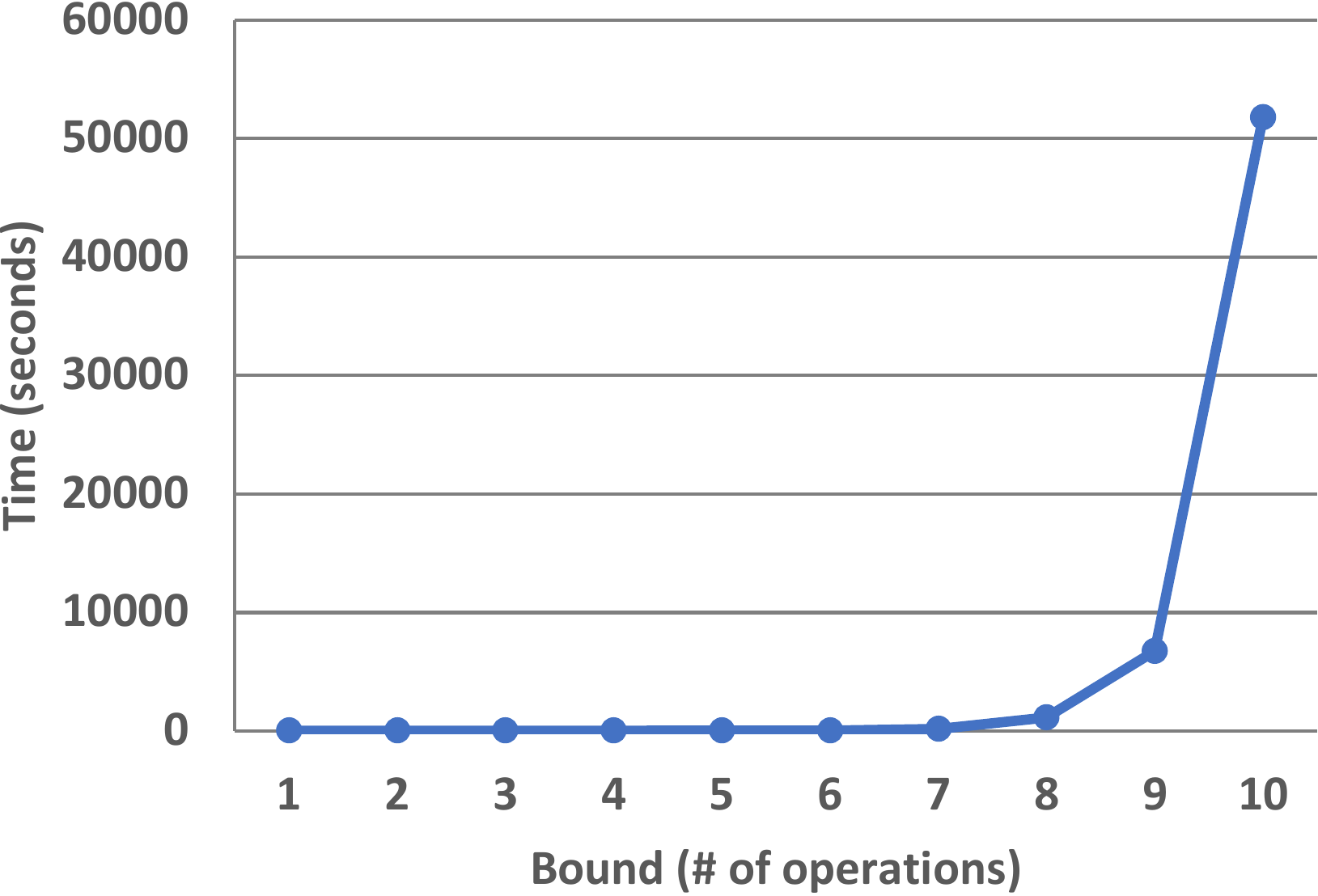}
  \vspace{-5pt}
  \caption{Runtimes for verifying \texttt{L1Hier} against \texttt{AtomicMemory} with varying bounds. Low bounds (e.g. 3-4) are sufficient to find common bugs.}
  \label{fig:int_verif_runtimes}
  \vspace{-10pt}
\end{figure}

\subsection{Methodology}
\label{sec:methodology}

RealityCheck is written primarily in Gallina, the functional programming language of Coq~\cite{coq:manual}. It also includes some OCaml, specifically the code to translate $\mu$spec++ formulae into Z3 using the Z3 OCaml API. 
RealityCheck builds on the Check suite's $\mu$spec parsing and axiom simplification~\cite{lustig:coatcheck}, adding support for the various $\mu$spec++ features discussed in Section~\ref{sec:uspecplusplus}.
In contrast to prior work~\cite{lustig:coatcheck} that used a basic SMT solver written in Gallina that supported $\mu$spec, RealityCheck utilises the state-of-the-art Z3 SMT solver~\cite{demoura:z3} to check its SMT formulae.
We extract our Gallina code to OCaml using Coq's built-in extraction functionality, and compile it and our OCaml code along with Z3's OCaml API into a standalone binary that can call into Z3 to solve SMT formulae.
Experiments were run on an Ubuntu 18.04 machine with 4 Intel Xeon Gold 6230 processors (80 total cores) and 1 TB of RAM. Each run of RealityCheck only utilises one core at time, but multiple instances of it can be run in parallel.

\subsection{Verifying Litmus Tests}
\label{sec:litmus_verif}

Table~\ref{tab:uarches} lists the 7 microarchitectures (each with 4 cores) on which we ran RealityCheck. The first six microarchitectures are comprised of the possible combinations of three pipelines and two memory hierarchies. The three pipelines are: (i) \texttt{inOrderCore}, an in-order 5-stage pipeline that performs memory operations in program order, (ii) \texttt{sbCore}, an in-order 5-stage pipeline with a store buffer from which it can forward values, and (iii) \texttt{rvwmoCore}, an out-of-order RISC-V pipeline which implements RISC-V's RVWMO weak memory model~\cite{riscv}. \texttt{sbCore} is capable of reordering writes with subsequent reads, as allowed by TSO (Total Store Order), the consistency model of Intel~\cite{intel:x86} and AMD x86 processors. \texttt{rvwmoCore} is more relaxed; it allows any loads and stores to different addresses to be reordered (in the absence of fences), while preserving address, data, and control dependencies as required by RVWMO. \texttt{rvwmoCore} supports coalescing of stores in its store buffer, and also supports all 8 of the RISC-V base ISA's fences for ordering memory loads and stores. The two memory hierarchies we model are \texttt{unifiedMem}, a single unified memory, and \texttt{L1Hier}, which consists of an L1 cache module backed by a module for a unified memory. The L1 cache in \texttt{L1Hier} is a non-blocking cache. It models requests for data from main memory, cache occupancy, and coalescing of stores before writing back to memory.

The final microarchitecture we evaluate RealityCheck on is \texttt{heteroProcRISCV}, a heterogeneous RISC-V processor. \texttt{heteroProcRISCV} has 2 out-of-order \texttt{rvwmoCore} pipelines and 2 in-order \texttt{sbCore} pipelines modified for the RISC-V ISA. To modify \texttt{sbCore} for RISC-V, we changed it to only enforce RISC-V fences that order writes with respect to subsequent reads, and to treat all other fences as nops. This is because the orderings enforced by other RISC-V fence types are already maintained by \texttt{sbCore} by default.

We also created four interfaces: one for each pipeline (\texttt{inOrderInt}, \texttt{sbInt}, and \texttt{rvwmoInt}) and one for the \texttt{L1Hier} memory hierarchy (\texttt{AtomicMemory}). Pipeline interfaces like \texttt{inOrderInt} reduce each pipeline module to its requests to and responses from memory (and its dependency orderings, in the case of \texttt{rvwmoInt}). Meanwhile, \texttt{AtomicMemory} abstracts \texttt{L1Hier} as a unified memory. These interfaces help verification scale as discussed below.

\subsubsection{SC and TSO Results}

Figure~\ref{fig:runtimes_sc_tso} shows RealityCheck's runtimes (as a box-and-whisker plot) for a suite of 95 SC/TSO litmus tests on the four SC and TSO microarchitectures, both with and without interfaces. These results use a bound of up to 11 operations per module. The litmus tests in the SC/TSO suite come from a variety of sources, including existing x86-TSO suites~\cite{owens:better} and automatically generated tests from the diy tool~\cite{diy}. As Figure~\ref{fig:runtimes_sc_tso} shows, RealityCheck's verification of litmus tests is quite fast, despite the increased detail of its microarchitectural specifications when compared to prior automated formal MCM verification. 92 of the 95 tests are verified by all 8 configurations in under 4 minutes each. The three remaining tests (\texttt{co-iriw}, \texttt{n3}, and \texttt{iwp27}) take longer as they have a large number of instructions (e.g. \texttt{n3} has 9 instructions) and/or possibilities to consider. However, RealityCheck still verifies them under all configurations in less than 14 minutes each.


Figure~\ref{fig:runtimes_sc_tso} also shows how the use of interfaces provides significant reductions in overall litmus test verification runtime. Pipeline interfaces and \texttt{AtomicMemory} can be used to abstract away portions of each design for litmus test verification. The use of abstraction reduces the total time to verify all litmus tests by 24.2\% for \texttt{cacheProc}, 0.6\% for \texttt{simpleProcTSO}, and 29.7\% for \texttt{cacheProcTSO}. Meanwhile, the use of abstraction increases runtime for \texttt{simpleProc} by 3.7\%, illustrating that interfaces may not reduce verification time for very simple designs. The runtime savings are much higher for \texttt{cacheProc} and \texttt{cacheProcTSO} because they use the \texttt{AtomicMemory} interface to abstract \texttt{L1Hier}. \texttt{L1Hier} is relatively detailed when compared to \texttt{AtomicMemory}, so verification using \texttt{AtomicMemory} takes much less time.

\subsubsection{RVWMO Results}

Figure~\ref{fig:runtimes_riscv} shows RealityCheck's runtimes for 98 RVWMO litmus tests on our 3 RVWMO microarchitectures, both with and without interfaces. These results use a bound of up to 11 operations per module. The RVWMO litmus tests used are generated using an automated litmus test synthesis tool~\cite{lustig:comprehensive}.
The litmus tests we use for RVWMO are the set of litmus tests up to 6 instructions long generated by this tool for the RVWMO MCM.

As Figure~\ref{fig:runtimes_riscv} shows, RealityCheck verifies each litmus test in all 6 configurations in under 4 minutes per test. The maximum time per test is lower under RVWMO than SC or TSO because our largest RVWMO litmus test is 6 instructions long (compared to e.g. the \texttt{n3} TSO litmus test which has 9 instructions). Similar to the SC and TSO microarchitectures, interfaces reduce verification time by 32.1\% for \texttt{cacheProcRISCV} and 30.0\% for \texttt{heteroProcRISCV}, while increasing runtime by 1.0\% for \texttt{simpleProcRISCV}.

The use of interfaces for abstraction depends on the implementations of those interfaces being verified against the interfaces. We present results on those next.

\subsection{Interface Verification}
\label{sec:interface_verif}


We conducted interface verification of three pipelines (\texttt{inOrderCore}, \texttt{sbCore}, and \texttt{rvwmoCore}) against their respective interfaces. Table~\ref{tab:int_verif} shows interface verification times for these pipelines (bound of 15). Interface verification of \texttt{rvwmoCore} takes longer than interface verification for the other two pipelines because the RISC-V pipeline is substantially more complicated. Nevertheless, its interface verification completes in 42 minutes.

\begin{table}[t]
\caption{Interface verification times for pipeline modules (bound of 15)}
\centering
    \begin{tabular}{ |c|c|c| } 
     \hline
     \textbf{inOrderCore} & \textbf{tsoCore} & \textbf{riscvCore} \\\hline 
     $<$ 1 sec. & 18 sec. & 42 minutes \\\hline
    \end{tabular}
    \vspace{-10pt}
  \label{tab:int_verif}
\end{table}

We also verified the \texttt{L1Hier} memory hierarchy against \texttt{AtomicMemory}. Figure~\ref{fig:int_verif_runtimes} shows interface verification runtime for \texttt{L1Hier} against \texttt{AtomicMemory} with varying bounds. \texttt{MemHier} interface verification runtimes at higher bounds are significantly larger than litmus test runtimes. For instance, interface verification of \texttt{L1Hier} at a bound of 10 takes over 14 hours. This is not surprising, as interface verification checks the implementation for all possible combinations of operations up to the user-specified bound. So for example, if verifying \texttt{L1Hier} against \texttt{AtomicMemory} with a bound of 10, one is checking all possible combinations of up to 10 transactions. This is essentially verifying all possible ``programs'' (from the perspective of memory) of up to 10 operations, which is far more than the possible memory transaction combinations that could result from a single litmus test.


The high runtimes for interface verification of \texttt{MemHier} are not as big an issue as they may initially seem. As Section~\ref{sec:bugfinding} below shows, bugs in implementations which cause them to not match their interfaces are detectable at lower bounds, and are found quickly even at higher bounds. Thus, even if interface verification has not terminated, if it does not find a bug quickly, the design is likely to be correct. Furthermore, interface verification can be run in parallel with both litmus test verification and with interface verification of other modules, making it well placed to take advantage of large compute clusters. Finally, interface verification of a module only needs to be run once, not once per litmus test. As the number of litmus tests run increases, the time saved from using interfaces for abstraction will draw closer to the time for interface verification.

\subsection{Bug Finding}
\label{sec:bugfinding}

To test how quickly interface verification can find bugs, we performed three case studies where we added a bug to the implementation of a component and then verified it against its interface. The first bug we added was to remove the axiom in \texttt{L1Hier}'s L1 cache which ensured that it could only have one value for a given address at any time. RealityCheck discovered this bug when verifying \texttt{L1Hier} against \texttt{AtomicMemory} at a bound of 3 in less than a second. Even when the bound was increased to 15, RealityCheck still found the bug in under 2 minutes. The second bug we added was to remove the axiom in \texttt{L1Hier}'s L1 cache which prevented it from dropping dirty values without writing them back. RealityCheck discovered this bug during interface verification at a bound of 4 in less than a second. Once again, even when the bound was increased to 15 operations, RealityCheck still caught the bug in less than 2 minutes. The final bug we added was to try and verify \texttt{sbCore} against \texttt{inOrderInt}. This should fail because \texttt{inOrderInt} requires program order to be preserved, while \texttt{sbCore} relaxes write-read ordering. RealityCheck duly discovered the bug at a bound of 15 in under a second.

\section{Related Work}
\label{sec:related}

There has been much work on formal MCM specifications of hardware ISAs in recent years~\cite{owens:better,sarkar:understanding,mador-haim:axiomatic,alglave:herding,zhang:gam,hower:heterogeneous,wickerson:rsp,gray:power,flur:arm,pulte:arm,trippel:tricheck,gaster:hsa,alglave:gpu}.
There has also been much work on MCM verification, using a variety of approaches. Fractal coherence~\cite{zhang:fractalcoh} uses a hierarchical design for coherence protocols to enable verifiability, but does not address MCM verification as a whole. Dynamic approaches like TSOTool~\cite{hangal:tsotool} and DVMC~\cite{meixner:dvmc} are incomplete in that they do not check all possible executions of the programs they are verifying. DVMC does modularise the system into the pipeline and coherence protocol, but no further. Furthermore, dynamic approaches require an implementation to exist. This is in contrast to RealityCheck, which can be used for early-stage design-time verification before RTL is written.

Automated formal approaches for hardware MCM verification consist of the Check suite~\cite{lustig:pipecheck,manerkar:ccicheck,lustig:coatcheck,trippel:tricheck,manerkar:rtlcheck} and PipeProof~\cite{manerkar:pipeproof}. The Check suite can automatically examine all possible executions of a litmus test on a microarchitecture to formally verify whether the test is observable on the microarchitectural specification. Meanwhile, PipeProof can automatically prove microarchitectural MCM correctness across all possible programs using an approach based on abstraction refinement. Despite their successes, the Check suite and PipeProof suffer from a lack of modularity that inhibits their ability to be fully usable in a real-world design setting. Their lack of modularity also encourages the writing of global axioms that do not directly match the underlying hardware. In contrast, RealityCheck supports modularity, hierarchy, and abstraction, which enable independent verification of each component, scalable verification of the entire design, and make it impossible to write global axioms. However, RealityCheck cannot guarantee correctness across all possible programs. Nevertheless, we believe that RealityCheck provides the basis for unbounded MCM correctness proofs of modular specifications in future work, just as litmus test-based verification approaches enabled the development of PipeProof~\cite{manerkar:pipeproof} for unbounded proofs of flat microarchitectural specifications.

The only prior hardware MCM verification work that supports modularity, hierarchy, and abstraction is Kami~\cite{murali:modular,choi:kami}. While Kami can prove correctness across all possible programs, these proofs must be written manually in a Coq framework, a setup which is not suitable for typical computer architects. In contrast, RealityCheck is an automated tool that is easy to use while still providing modularity, hierarchy, and abstraction, though it cannot prove correctness across all possible programs.

Coppelia~\cite{zhang:coppelia} searches for hardware security exploits by generating a C++ class hierarchy from processor Verilog and conducting backward symbolic execution on it. However, unlike RealityCheck, it evaluates the entire processor at once, and it does not conduct MCM verification.
\section{Conclusion}
\label{sec:conclusion}

Modern processors are complex parallel systems which incorporate components built by many different teams, and they require stringent MCM verification to ensure their correctness.
In order to work with the distributed nature of the hardware design process, specification and verification of orderings enforced by hardware components would ideally be modular. However, all prior automated formal hardware MCM verification approaches provide no support for modularity, making it hard for them to be used in real-world design settings.

In this paper we present RealityCheck, the first methodology and tool for automated formal MCM verification of modular hardware design specifications. RealityCheck provides support in its novel specification language $\mu$spec++ for modularity, hierarchy, and abstraction. This allows users to encapsulate orderings enforced by a component into modules which can be composed with each other to create larger modules. Each component can be verified against its interface specification independently of the rest of the system, and regardless of whether the rest of the design specification exists. Verification of the entire design can be split into multiple smaller verification problems, allowing verification to scale. We implement RealityCheck as an automated tool and show that it is capable of verifying hardware designs across a range of litmus tests, and that it can detect bugs extremely quickly. In summary, RealityCheck is a significant step forward on the road to fully verifying today's heterogeneously parallel SoCs.


\bibliographystyle{IEEEtranS}
\bibliography{refs}

\end{document}